\documentclass[12pt]{article}
\usepackage{amsfonts,amssymb,amsmath,graphicx,hyperref}

\title{Trajectories and  Particle Creation and Annihilation in Quantum
Field Theory}
\author{
Detlef D\"urr\footnote{Mathematisches Institut der Universit\"{a}t 
	M\"{u}nchen, Theresienstra{\ss}e 39, 80333 M\"{u}nchen, Germany. 
	E-mail: duerr@mathematik.uni-muenchen.de},
Sheldon Goldstein\footnote{Department of Mathematics -
	Hill Center, Rutgers, The State University of New Jersey, 
	110 Frelinghuysen Road, Piscataway, NJ 08854-8019, USA.
	E-mail: oldstein@math.rutgers.edu},\\
Roderich Tumulka\footnote{Mathematisches Institut der Universit\"{a}t 
	M\"{u}nchen, Theresienstra{\ss}e 39, 80333 M\"{u}nchen, Germany. 
	E-mail: tumulka@mathematik.uni-muenchen.de},\ and
Nino Zangh\`\i\footnote{Dipartimento di Fisica, INFN sezione di Genova, 
	Via Dodecaneso 33, 16146 Genova, Italy. E-mail: zanghi@ge.infn.it}
}
\date{}

\begin{document}

\maketitle
\begin{abstract}
We develop a theory based on Bohmian mechanics in which particle world
lines can begin and end. Such a theory provides a realist description
of creation and annihilation events and thus a further step towards a
``beable-based'' formulation of quantum field theory, as opposed to
the usual ``observable-based'' formulation which is plagued by the
conceptual difficulties---like the measurement problem---of quantum
mechanics.
\bigskip

\noindent PACS numbers \\
	03.65.Ta (foundations of quantum mechanics)\\
	42.50.Ct (quantum description of interaction of light and matter)
\end{abstract}
\hyphenation{equi-variant}

\newcommand{\RRR}{\mathbb{R}} %
\newcommand{\CCC}{\mathbb{C}}
\newcommand{\SSS}{\mathbb{S}} %
\newcommand{\EEE}{\mathbb{E}}
\newcommand{\NNN}{\mathbb{N}}
\newcommand{\bo}{\mathrm{B}}
\newcommand{\el}{\mathrm{F}} 
\newcommand{\inter}{\mathrm{int}} 
\newcommand{\profile}{\varphi} 
\newcommand{\hi}{${\cal H}$} 
\newcommand{\LL}{{\cal L}}
\newcommand{\Fock}{{\cal F}} 
\newcommand{\1}{\boldsymbol{1}} 
\newcommand{\vx}{\boldsymbol{x}}
\newcommand{\vk}{\boldsymbol k}
\newcommand{\vu}{\boldsymbol u}
\newcommand{\vy}{\boldsymbol y}
\newcommand{\vX}{\boldsymbol X}
\newcommand{\vY}{\boldsymbol Y}
\renewcommand{\Im}{\,\mathrm{Im}\,}
\renewcommand{\Re}{\,\mathrm{Re}\,}
\newcommand{\Laplace}{\Delta} 
\newcommand{\D}{d} 
\newcommand{\E}{e} 
\newcommand{\I}{\mathrm{i}} 

\section{Introduction}

A formulation of nonrelativistic quantum mechanics based on objectively
existing particle positions and particle trajectories, now usually called
Bohmian mechanics, was proposed fifty years ago by David~Bohm (and even
earlier by Louis de Broglie); see \cite{BM1} for a recent overview.  Today
there remain two big challenges for this approach: to form a relativistic
version, and a version suitable for quantum field theory (QFT). Here we
shall address the latter. We describe a general and particularly natural
way of extending Bohmian mechanics to QFTs, explicitly giving the equations
for a simple example.

Bohm himself proposed \cite[p.~230]{BH} that Bohmian mechanics should be
extended to QFT by means of the incorporation of the actual field
configuration, guided by a wave functional (the state vector). In contrast,
John~Bell proposed a model \cite[p.~173]{BellBible} in which, instead of
the field configuration, the local beables are the fermion numbers at each
site of a lattice discretizing 3-space.  We argue that it is instructive to
modify Bell's proposal in two ways, and thus get a similar but even
simpler theory with a direct connection to Bohmian mechanics. Below, we
give an explicit example of such a theory for a particularly simple
Hamiltonian.

Bell's model contains no beables representing the bosonic degrees of
freedom (such as radiation), neither an actual field, nor actual particles,
nor anything else; the existence of a radiation part of the state vector
is relevant  only to the behavior of the fermions. This is certainly
consistent and empirically irrefutable, but it is neither
\emph{necessary} (as our example below shows), nor even a
\emph{natural} view.\footnote{
After all, if the role of the wave function is to guide particles, it would
seem that there should be as many particles as there are variables in the
wave function. In addition, the Hilbert space of the radiation degrees of
freedom (unlike the one for quark color) is not a purely abstract Hilbert
space, but is related to space-time points (via creation and annihilation
operators, for example), a fact that would seem surprising if the state
vector were not related to space-time objects such as particles, strings,
or fields.
} In the model proposed below, the bosons have the same status as the
fermions: they are particles, described by their positions.  This, of
course, should not be regarded as discouraging consideration of the
approach based on actual field configurations.

The other deviation from Bell's proposal is the replacement of the lattice
by continuous space.  The lattice was introduced in the first place for the
purpose of providing an effective ultraviolet cut-off and thus a
well-defined Hamiltonian. Such a cut-off, however, can also be realized by
smearing out the interaction Hamiltonian through convolution with, say, a
sharply peaked but bounded function $\profile$. The continuum analogue of the particle
number at every lattice site is the position of all particles in ordinary
space, with the total number of particles $N(t)$ possibly varying with
time. However, in the model we propose the particles follow Bohmian
trajectories, except when there is particle creation or annihilation. (For a
discussion of how Bohmian mechanics arises from a lattice model (in the
absence of interaction) in the limit of vanishing lattice width, see
\cite{Sudbery,Vink}.)

Our proposal profits, we believe, from making this contact with
Bohmian mechanics, since then every argument for taking the Bohmian
trajectories seriously also provides some support for the
proposal. Bohmian mechanics also profits from this contact because the
Bohmian trajectories \emph{can} then be taken seriously even in the
framework of QFT.

Our moving configuration $Q(t)$ is constructed in such a way that it is
random with distribution at time $t$ equal to $\rho(t)=|\Psi(t)|^2$, where
$\Psi(t)$ is the (position) Fock space representation of the state vector
$|\Psi(t)\rangle$ at time $t$. (For a model in which there are only bosons,
the $n$-particle component of $\Psi$ is
\begin{equation}\label{Fockfunction}
  \Psi^{(n)}(\vx_1,\ldots,\vx_n) = \frac{1}{\sqrt{n!}} \, \langle 0 |
  a(\vx_1)\cdots a(\vx_n) | \Psi\rangle 
\end{equation}
where $|0\rangle$ is the Fock vacuum and $a(\vx) = (2\pi)^{-3/2} \int\!\D^3
\vk\, \E^{\I\vk\cdot\vx} \, a_{\vk}$ is the boson annihilation operator at
position $\vx$.)  In particular, the probability that $N(t)=n$, i.e., that
there are $n$ particles at time $t$, equals the integral of $\rho(t)$ over
$K^{(n)}\cong (\RRR^3)^n$ (suitably symmetrized), the $n$-particle sector
of configuration space---i.e., the $L^2$ norm of the projection of
$|\Psi\rangle$ onto the $n$-particle subspace of Fock space. A nontrivial
superposition of quantum states with different particle numbers leads to a
probability distribution over different particle numbers, among which only
one is, of course, actually realized.

\section{Configuration Jumps}

The Hamiltonian for QFTs is typically a sum of terms, each of which yields
a contribution to the motion we wish to propose. It is quite generally the
case that $H=H_0+H_{\inter}$ where the free Hamiltonian $H_0$ corresponds
naturally to a deterministic motion---in the model considered here that of
Bohmian mechanics, and in a relativistic model for example the Bohm--Dirac
motion \cite[p.~274]{BH}---given by a (time-dependent) velocity field
$v=v^{\Psi(t)}(q,t)$, on configuration space $K=\cup_n K^{(n)}$, that
defines the ``deterministic part'' of the process, while the interaction
term $H_{\inter}$ corresponds to random jumps, a continuum version of the
process proposed by Bell \cite[p.~173]{BellBible}, defined by jump rates
$\sigma=\sigma(q',q,t)=\sigma^{\Psi(t)}(q',q)$ for a transition from $q$ to
$q'$ at time $t$. [This means that when the actual configuration $Q$ at
time $t$ is $q$, then with probability (density, with respect to \ $q'$) 
$\sigma(q',q,t) \,\D t$, $Q$
will jump from (very near) $q$ to $q'$ in the time-interval
$(t,t+\D t)$.]\footnote{One can regard the jumps as having two ingredients: A
jump occurs with total rate $\bar\sigma(q,t)=\int \sigma(q',q,t)\,dq'$;
when a jump does occur, the destination is randomly chosen with
distribution $\sigma(q',q,t)/\bar\sigma(q,t)$.}

The relevant continuum analogue of Bell's jump rates for $H=H_{\inter}$,
equations (6--8) of \cite[p.~173]{BellBible}, is
\begin{equation}\label{sigma}
  \sigma^{\Psi}(q',q)= {\frac2{\hbar}}\frac{\left(-\Im
  \overline{\Psi(q)}\langle q| H_{\inter} | q' \rangle
  \Psi(q')\right)^+}{\overline{\Psi(q)}\Psi(q)} 
\end{equation}
where for particles with spin the two products in the numerator and the
product in the denominator are local spinor inner products, and where we
have used the notation $A^+ = \max(A,0)$ for the positive part of
$A\in\RRR$. This will typically be well-defined (as jump rates), since the
kernel $\langle q| H_{\inter} | q' \rangle$ of the (cut-off) interaction
Hamiltonian of a QFT should involve nothing worse than $\delta$-function
singularities.

The complete process, corresponding to the total Hamiltonian, is then given
by the deterministic motion with velocity $v$, randomly interrupted by
jumps, with rate $\sigma$, after each of which the deterministic motion is
resumed until it is again interrupted. As a function of $t$, each
realization $Q(t)$ is thus piecewise smooth. At the end of a smooth piece,
$Q$ jumps to the starting point of the next smooth piece.  What is
stochastic about $Q$ are the times at which the jumps take place, and the
destinations of the jumps. The probabilities for times and destinations are
governed by the wave function. $Q(t)$ is a Markov process. 

As described, our process is so designed as to directly imply the
following equivariance theorem: \textit{If $Q(t_0)$ is chosen at
random with distribution $|\Psi(t_0)|^2$, then at every later time
$t>t_0$, $Q(t)$ is distributed with density $|\Psi(t)|^2$.} (This can
also be explicitly checked by comparing the equation for $\partial
(\overline{\Psi} \Psi)/\partial t$ as implied by the evolution
equation of the quantum state, $i\hbar\,\partial\Psi/\partial t = H
\Psi$, and the master equation for the distribution of $Q(t)$, which
reads
$$
  \frac{\partial}{\partial t} \rho(q,t) = 
  -\nabla\cdot \Big( \rho(q,t)\, v(q,t)\Big) +
  \int_K \! \D q' \big( \rho(q',t)\,\sigma(q,q',t) -
  \rho(q,t)\,\sigma(q',q,t) \big) \,.\ )
$$

The choice of jump rates that will make the $|\Psi|^2$ distribution
equivariant is not unique. But the `minimal' jump rates are unique, and
these are the ones we have chosen. Details on the choice of jump rates as
well as an elucidation of the general mathematical structure underlying
equivariant processes will be presented in \cite{creation2}.

Corresponding to the standard interaction terms in QFT, the possible jumps
are very restricted: there are only changes in the particle number by $\pm 1$,
and possible types are (i) appearance or (ii) disappearance of a particle
(while the others remain at their positions) or (iii) replacement of a
particle by two others or (iv) the reverse of this. Type (i) is
appropriate for, e.g., photon emission, (ii) for absorption, (iii) for pair
creation and (iv) for pair annihilation. (It follows that there is no
discontinuity in the individual particle world lines, in spite of the
discontinuity in $Q$.)  In our explicit model, only types (i) and (ii)
occur.

\section{An Explicit Model}

We now present an explicit theory. It is based on a ``baby'' QFT taken
from \cite[p.~339]{Schweber} and \cite{Nelson}, containing two species
of particles, which we simply call electrons and photons. Electrons,
whose number stays constant, emit and absorb photons. To keep things
simple, we employ nonrelativistic dispersion relations (as everything
here is nonrelativistic) for both electrons and photons, and give the
photon a positive (rest) mass. In addition, we ignore spin and
polarization, and, of course, smoothen the interaction Hamiltonian.

In field theoretic language, we have a bosonic field $\phi(\vx) =
a^{\dag}(\vx) + a(\vx)$, with $a^{\dag}$ and $a$ the photon creation
and annihilation operators [cf. equations (\ref{cr}) and (\ref{an})]
and a fermionic field $\psi(\vx)$, and the Hamiltonian is the sum
\begin{eqnarray}\nonumber
  H&=&H_{F}\ + H_{B}\ +\ H_\inter \\\nonumber
  &=&(1/2m_F) \int\!\D^3\vx\,\nabla \psi^{\dag}(\vx)\nabla \psi(\vx)
  +(1/2m_B) \int\!\D^3\vx\,\nabla a^{\dag}(\vx)\nabla a(\vx)\\\nonumber 
  &\quad&\qquad + g\int \!\D^3\vx\, \psi^{\dag}(\vx) \phi_{\profile}
  (\vx) \psi(\vx)
\end{eqnarray}
where $\phi_{\profile}(\vx)= \int\!\D^3\vy\big(\profile(\vx-\vy)
a^{\dag}(\vy) + \overline{\profile}(\vx-\vy)a(\vy)\big)$ is the cutoff
bosonic field, $m_\el$ and $m_\bo$ denote the mass of the electrons
and photons, and $g$ is a real coupling constant.  $H$ commutes with
the fermion number operator $N_F =\int\!\D^3\vx\,
\psi^{\dag}(\vx)\psi(\vx)$.

Since the fermion number is conserved, we give it a fixed value
$N$. The configuration space (changing notation slightly from before) is $K =
\bigcup_{m=0}^\infty K^{(m)}$ where $m$ is the photon number and
$$
  K^{(m)} := (\RRR^{3})^N \times (\RRR^{3})^m\,, \quad
  K^{(0)} = (\RRR^3)^N\,.
$$
We will denote the electron coordinates by
$x:=x^{(N)}:=(\vx^1,\ldots,\vx^N)$  and the photon coordinates by
$y:=y^{(m)}=(\vy^1,\ldots,\vy^m)$. The full configuration is thus given by
$q=(x,y)$ and, more explicitly, by $q^{(m)}=(x,y^{(m)})$. 

In terms of the wave function $\Psi$ [the (position) Fock representation of 
the quantum state], the Hilbert space inner product is
$$
  \langle \Phi|\Psi\rangle := \sum_{m=0}^\infty \int_{K^{(m)}}\!\!\!
  \D^{3N}x\, \D^{3m}y\,\, \overline{\Phi}(x,y)\,
  \Psi(x,y)\,,
$$
and the contributions to the Hamiltonian assume the form $$H_\el = -\sum_i
\frac{\hbar^2}{2m_\el} \Laplace_i$$ $$H_\bo = -\sum_j
\frac{\hbar^2}{2m_\bo} \Laplace_j$$  and
\begin{equation}\label{HamiltonianInter}
  H_\inter = g \sum_{i=1}^N \phi_\profile(\vx^i)=
  g \sum_{i=1}^N\left( a^{\dag}_\profile(\vx^i) +
  a_{\overline\profile}(\vx^i)\right) 
\end{equation}
with creation and annihilation operators $a^\dag_\profile$ and
$a_{\overline\profile}$ acting on Fock space in a smeared-out form:
\begin{eqnarray}\nonumber
  a^{\dag}_\profile(\vx)
  &=&
  \int\!\D^3\vu \,\profile(\vu-\vx)\, a^{\dag}(\vu)
  \\\nonumber
  a_{\overline\profile}(\vx)
  &=&
  \int\!\D^3\vu \,{\overline\profile}(\vu-\vx)\, a(\vu),
\end{eqnarray}
where
\begin{eqnarray}\label{cr}
  \big(a^\dag (\vu)\Psi\big)(q^{(m)}) 
  &=&
  \frac{1}{\sqrt{m}} \sum_j\delta(\vy^j-\vu) \,
  \Psi(\widehat{q^j}) \,,
  \\\label{an}
  \big(a(\vu) \Psi\big)(q^{(m)}) 
  &=&
  \sqrt{m+1} \, \Psi\big(q^{(m)},\vu\big).
\end{eqnarray}
$a^\dag_\profile(\vx)$ creates a new photon in state $\profile$ centered at
$\vx\in \mathbb{R}^3$ (which will be the position of an electron), and
$a_{\overline\profile}(\vx)$ annihilates a photon with ``form factor''
$\profile$ at $\vx$.\footnote{If the form factor $\profile$ is
square-integrable (as we assume) it provides an ultraviolet cutoff; it can
be regarded as determining the effective range of a electron's power to
create or absorb a photon.} Here $(q^{(m)},\vu)$ is the configuration with a
photon at $\vu$ added to $q^{(m)}$ and $\widehat{q^j}$ is the configuration
with the $j$-th photon deleted from $q^{(m)}$. Thus
\begin{eqnarray}\label{Schroedinger}
  (H_{\inter}\Psi)(q^{(m)})
&=& 
  \frac{g}{\sqrt{m}} \sum_{i=1}^N \sum_{j=1}^m \profile(\vy^j-\vx^i)
  \, \Psi \big(\widehat{q^j}\big) \, 
  \\\nonumber
  &+&
  g\sqrt{m+1} \sum_{i=1}^N  \int \!\D^3 \vy'\,
  \overline\profile (\vy'-\vx^i) \, \Psi \big(q^{(m)},\vy'\big)\ .
\end{eqnarray}
$\Psi$ satisfies the Pauli principle, i.e., it is symmetric in the photon
variables and antisymmetric in the electron variables.

We now turn to the particles, described by the actual electron configuration $X$
and the actual photon configuration $Y$.   The deterministic part of the
motion $Q(t)=(X(t),Y(t))$,
corresponding to $H_{\el}+ H_{\bo}$, is the usual Bohm motion
\begin{eqnarray}
 \label{EqMotionA}
  \dot{X}^{i} 
  &=&
  v_{\el,i}^{\Psi}(Q)\; := \;\frac{\hbar}{m_\el} \Im
  \frac{\big(\partial \Psi/\partial \vx^{i}\big)
  (Q)}{\Psi(Q)}\,,
  \\
 \label{EqMotionB}
  \dot{Y}^{j} 
  &=&
  v_{\bo,j}^{\Psi}(Q)\; :=
  \;\frac{\hbar}{m_\bo} \Im
  \frac{\big(\partial \Psi/\partial \vy^{j}\big)
  (Q)}{\Psi(Q)}
\end{eqnarray}
and it follows from  (\ref{sigma}) and (\ref{Schroedinger}) that only two kinds of
jumps occur, with rates as follows: 
\begin{itemize}
\item The $j$-th photon vanishes, while all the other particles stay at
their positions. Thus $q'=\widehat{q^j}$ and the jump rate from
$q=q^{(m)}$ to $q'$ is
\begin{equation}\label{annrate}
\sigma(q',q) =   \frac{2g}{\hbar\sqrt{m}} \Big[ -\sum_{i=1}^N \Im 
  \frac{\Psi(\widehat{q^j}) \,
  \profile(\vy^j-\vx^i)}{\Psi(q)} \Big]^+.
\end{equation}

\item A new photon appears at  $\vy'$, while all the other particles stay at
their positions. Thus $q'=(q^{(m)},\vy')$ and the jump rate from
$q=q^{(m)}$ to $q'$ (a density in the variable $\vy'$) is 

\begin{equation}\label{destinationdistribution}
\sigma(q',q) =   \frac{2g}{\hbar}\sqrt{m+1}
  \Big[ -\sum_{i=1}^N \Im \frac{\Psi(q,\vy')
  \,\overline\profile (\vy'-\vx^i)}{\Psi(q)} \Big]^+.
\end{equation}
\end{itemize}
We reiterate that these choices guarantee that the process obeys the
equivariance theorem mentioned in the previous section.

Note that, contrary to what might have been expected, the creation
operator in the interaction Hamiltonian (\ref{HamiltonianInter})
corresponds, according to (\ref{sigma}), to the annihilation rate,
while the annihilation operator corresponds to the creation
rate. There is less in this than meets the eye: The correspondence is
an artifact of the way (\ref{sigma}) is written, and if (\ref{sigma})
had been written equivalently as
\begin{equation}
  \sigma^{\Psi}(q',q)= {\frac2{\hbar}}\frac{\left(\Im
  \overline{\Psi(q')}\langle q'| H_{\inter} | q \rangle
  \Psi(q)\right)^+}{\overline{\Psi(q)}\Psi(q)} 
\end{equation}
the correspondence would have been reversed. Note 
also that if $\profile$ is supported by a $\delta$-ball around the origin
of $\RRR^3$, then photons can be created or annihilated only in the
$\delta$-neighborhood of an electron.\footnote {
One might suspect that if $\delta$ were small, there would be only a small
probability (perhaps of order $\delta^3$) that a photon will come closer
than $\delta$ to an electron. But equivariance implies that this is not so:
if a certain amount of $|\Psi^{(m)}|^2$ flows to the sector $K^{(m-1)}$,
then the probability for a photon to be annihilated  is just as large.
}

Our model is translation invariant (unless one introduces an
additional external potential), time translation invariant, rotation
invariant provided $\profile$ is spherically symmetric, time reversal
invariant provided $\profile$ is real-valued, and gauge invariant
provided one introduces an external vector potential---in all
derivatives acting on electron coordinates---and an external scalar
potential into $H_\el$. Galilean boost invariance fails, but not
because of the Bohmian variables; owing to the interaction term
(\ref{Schroedinger}), it fails, in fact, for the evolution of the
quantum state. The reason is that, roughly speaking, a photon gets
created with wave function $\profile$ which cannot be Galilean
invariant. Alternatively, one could say that under boosts with
velocity $\vu$ the form factor $\profile(\vy-\vx)$ must be replaced by
$\exp{(\I m_\bo \vu\cdot \vy/\hbar)}\, \profile(\vy-\vx)$. Note that
even with the cutoff removed, i.e., for $\profile(\vy)=\delta(\vy)$,
the quantum dynamics is not Galilean invariant.

If we add to $H$ a suitable confining potential (and $\profile$ is
real-valued, and the rest energy of the photon is made positive), it
possesses a unique ground state \cite{Spohn}, and this ground state (like
every nondegenerate eigenstate) is real up to an overall phase. Thus in
this state the jump rates (\ref{annrate}), (\ref{destinationdistribution})
as well as the velocities (\ref{EqMotionA}), (\ref{EqMotionB}) vanish
identically---nothing moves. Surprisingly, perhaps, the Bohmian particles
do not perform any ``vacuum fluctuations''.

As in Bohmian mechanics, disentangled subsystems are governed by the same
laws as for the whole. More precisely, if $\Phi$ and $\Psi$ are two wave
functions from Fock space having disjoint supports $S_\Phi$ and $S_\Psi$ in
physical space $\RRR^3$, $\Phi\otimes\Psi$ defines another Fock state $\Phi
\odot \Psi$ after suitable symmetrization; if the distance between $S_\Phi$
and $S_\Psi$ is at least the diameter $\delta$ of the support of
$\profile$, then in a system with wave function $\Phi\odot \Psi$ the
particles in $S_\Phi$ and those in $S_\Psi$ will not influence each other,
each set moving independently according to the corresponding
`factors'. This is a consequence of the fact that the velocities and jump
rates are homogeneous of degree 0 in the wave function.

While the other particles keep their positions when a photon is created or
annihilated, their velocities may change discontinuously because
(\ref{EqMotionA}) evaluated at the destination may differ from
(\ref{EqMotionA}) evaluated at the point of departure. As a result, the
world lines of all (possibly distant) particles, if the particles are
entangled, will have kinks at the times of particle creation or
annihilation. These kinks will however not be visible in, say, a cloud
chamber since the necessary entanglement is destroyed by the decoherence
of the tracked particle with its environment, caused, say, by the particle's
interaction with the vapor.

\section{Removing the Cutoff}

Removing the cutoff is of course problematical, which is why the cutoff was
introduced in the first place. However the problems arise primarily from
the evolution equation of the wave function, not those of the Bohmian
configuration, equations (\ref{EqMotionA}) through
(\ref{destinationdistribution}). Suppose that a family of $\profile$'s is
parametrized by $\Lambda\in\RRR$, and that as $\Lambda\to\infty$,
$\profile^\Lambda(y)$ approaches $\delta(y)$.  Then the limit
$\Lambda\to\infty$ corresponds to ``removing the cutoff''. Unfortunately,
the corresponding $H^\Lambda$ will not converge in any reasonable
sense. But there exist numbers $E^\Lambda$ tending to infinity
\cite{Nelson} such that $H^\Lambda - E^\Lambda$ does converge in a suitable
sense, and the evolution of the wave function is well-defined. This seems
completely acceptable. We do not know whether the corresponding process
$Q^\Lambda$ approaches (given a fixed initial wave function) a limiting
process $Q^\infty$. To decide this requires a careful mathematical study,
but at least we see nothing precluding this possibility: the velocity law 
is not affected by the cutoff, while absorption might become deterministic 
in the limit $\Lambda\to\infty$ and occur whenever (and only when) a photon
hits an electron.

Other Hamiltonians, more sophisticated ones, are more problematical, and do
not possess a limit as $\Lambda \to \infty$, even after subtracting an
``infinite energy''. In some cases this is due to the creation of a large
(average) number $m^\Lambda$ of photons that goes to infinity with
$\Lambda$. On the other hand, it is not clear that removing the cutoff is
desirable or relevant. That is, there might exist an effective UV cutoff in
nature, just as there is an effective IR cutoff (the finite radius of the
universe).

Be that as it may, if the unitary evolution on some Hilbert space does not
survive the limit $\Lambda\to\infty$, we face a problem, one that seems
particularly bad for Bohmian theories, which so heavily rely on the wave
function and, consequently, its having a well-defined unitary
evolution. But appearances are misleading here. From a Bohmian viewpoint,
the basic variable, bearing all the physical implications of the theory, is
the configuration $Q$, whereas $\Psi$ and $H$ are only theoretical objects
whose purpose is to generate a law of motion for $Q$. And, indeed, a law
for $Q$ might arise as a limit $\Lambda\to\infty$ of the law induced by
$\Psi^\Lambda$ and $H^\Lambda$; it might be the case that while
$\Psi^\Lambda$ and $H^\Lambda$ do not have a limit, the time evolution of
$Q$ is well defined in the limit. After all, this is precisely what occurs
when one consider the limit $\hbar\to 0$ of nonrelativistic quantum
mechanics: while the Hamiltonian and wave function do not have any sensible
limit, the law for $Q$ does!

\section{Determinism}

We close with a remark on (the lack of) determinism. It may seem surprising
that we abandon determinism. Was not the main point of hidden variables to
restore it? Actually, no. What was important was to provide a clear and
coherent account of quantum mechanics. The simplest such account, we
believe, is provided by Bohmian mechanics, which happens to be
deterministic. And the simplest such account of QFT seems to be of the sort
we have presented here, which is stochastic.

\end{document}